\documentclass[runningheads]{llncs}

\usepackage[T1]{fontenc}

\usepackage{graphicx}

\usepackage{color}

\PassOptionsToPackage{dvipsnames}{xcolor}
\usepackage{xcolor}

\usepackage{amsmath}
\usepackage{hyperref}
\usepackage{listings}
\usepackage{tcolorbox}
\usepackage{pgf, pgffor}
\usepackage{pdfpages}
\usepackage{xspace}
\usepackage{booktabs, multirow}
\usepackage{soul}
\usepackage{wrapfig}

\usepackage{tikz}
\usetikzlibrary{shapes.geometric, arrows, calc}
\usepackage{amssymb}
\usepackage{array}       
\usepackage{subcaption}

\usepackage{svg}

\begin{document}

\title{Can LLMs Enable Verification in Mainstream Programming?}
%
%

\author{}
\institute{}
\authorrunning{}

\author{Aleksandr Shefer\inst{1,2} \and 
Igor Engel\inst{1,2} \and 
Stanislav Alekseev\inst{1,3} \and 
Daniil Berezun\inst{1} \and 
Ekaterina Verbitskaia\inst{1,2} \and 
Anton Podkopaev\inst{1,2} } 
\authorrunning{A. Shefer et al.}
%
\institute{JetBrains Research, Amsterdam, the Netherlands \and 
Constructor University, Bremen, Germany \and 
Neapolis University, Pafos, Cyprus }
%
\maketitle              

\newcommand{\todo}[1]{{#1}}

\renewcommand\UrlFont{\color{blue}\rmfamily}
\urlstyle{rm}

\begin{abstract}

Although formal methods are capable of producing reliable software, they have seen minimal adoption in everyday programming. 
Automatic code generation using large language models is becoming increasingly widespread, but it rarely considers producing strong correctness guarantees.
In this study, we explore the ability of LLMs to produce verified code in three verification languages  (Dafny, Nagini, and Verus). 
To do so, we use manually curated datasets derived from the state-of-the-art Python benchmark, HumanEval. 
We also assess what types of information are sufficient to achieve good-quality results. 

\end{abstract}
\section{Introduction}


The cost of software failures only in the US is estimated to reach trillions of dollars a year~\cite{krasner2022cost}. 
Ever since the advent of large language models, a multitude of code generators have been developed. 
With people slowly beginning to trust these systems, significantly less time is spent critically evaluating the produced code. 
As a result, errors are more likely to find their way into codebases, leading to unintended consequences.
Thus, it is of paramount importance to ensure the correctness of the generated code. 

Formal methods help programmers prevent avoidable mistakes by offering tools to analyse and prove the correctness of their code. 
It is most well-adopted in the areas where a mistake can have severe consequences, such as cryptography~\cite{zinzindohoue2017hacl}, finance~\cite{cassez2022formal}, and aerospace~\cite{souyris2014industrial}. 
Unfortunately, verification requires extensive expertise and extends the time necessary to create programs, which prevents its use in less safety-critical domains. 

Systems such as Dafny and F* automate proof search through the use of Satisfiability Modulo Theories (SMT) solvers, lowering the amount of effort needed for verification of complex properties. 
However, these verifiers feature specialized programming languages for both implementation and proofs. 
In order to introduce certified code into a software project, one has to commit to using a new language, which frequently entails less mature developer environments, limited interoperability with mainstream languages, and a steeper learning curve for the engineers. 

A promising solution to this problem comes in a form of intermediate verification languages such as Viper~\cite{muller2016viper}. 
With this approach, an algorithm can be implemented in a restricted subset of a popular programming language directly and then supplemented with formal specification and proofs. 
This helps bridge the gap between mainstream programming and formal methods, reducing the barriers for adoption. 
The key challenge is then ensuring that reasoning about the program remains as straightforward as possible. 

A lot of effort has been put into solving this challenge over the years. 
There are many obstacles to the problem of automated software verification and synthesis.
The most crucial stems from the undecidability of automatic loop invariant synthesis.
There are many classical approaches to tackle the problem, including 
    dynamic detection of likely invariants~\cite{le2019sling,ernst2007daikon,boshernitsan2006daikon},
    inductive invariant synthesis~\cite{riley2022multi,beyer2007path,dillig2013inductive},
    data-driven approaches~\cite{lu2020novel,lingsound,hu2015automatic}, 
    deductive synthesis~\cite{young2024higher},
    and non-linear reasoning~\cite{kincaid2017non,srikanth2017complexity}.
Most state-of-the-art tools, which provide any guarantees of soundness and correctness, rely on SMT-solvers to infer invariants for recursion and iteration.
This is notably true for intermediate verification languages like Viper and Verus, as well as F* and Dafny.
In this research, we do not focus on improving the underlined theory neither do we attempt to refine solvers' results, leaving this important task to the designers and researches of SMT-solvers and verification languages.
Instead, we aim to determine the extent to which modern LLMs are capable of bridging the existing gap and generate invariants for these languages that can be successfully verified by solvers.

Generative AI solutions have been able to achieve rather high success rates in producing partial~\cite{mugnier2024laurel,silva2024leveraging,kamath2023finding} and complete~\cite{yao2023leveraging,poesia2024dafny,chen2024automated} proofs for existing code, as well as inferring postconditions based on textual descriptions~\cite{endres2024can} and complete verified code in Dafny~\cite{sun2024clover}, Rust~\cite{aggarwal2024alphaverus}, and F*~\cite{chakraborty2024towards}. 
The ability of large language models to make progress towards automating formal verification for specialized systems suggests it may be possible to integrate formal methods into mainstream programming. 

In this paper, we explore the capability of LLMs in generating formally verified code across three different systems. 
Our focus lies on Nagini~\cite{eilers2018nagini} and Verus~\cite{lattuada2023verus}~--- the verifiers of subsets of popular programming languages Python and Rust.
We also include Dafny into consideration, which serves as a baseline. 
In addition to this, we investigate what kind of problem specification is enough for the model to produce verified code by exposing different amounts of information to it. 
Finally, we prepare three datasets based on HumanEval~\cite{chen2021evaluating} of high-quality, manually implemented, verified programs to function as reference solutions. 
\section{Method}

Our overarching goal of simplifying the lives of programmers who may not be used to formal verification poses the following question: how much information do we have to expose to an LLM in order to successfully produce correct-by-construction code? 
Can AI only help with the inferring simple loop invariants and assertions to finish the proof based on the existing codebase? 
Would it be possible to also generate implementations and high-level specifications from a natural language description? 
As we have found out, the answer lies somewhere in between. 
In pursuit of this research goal, we designed a universal framework that prepares prompts and validates programs generated in several scenarios. 
In this section, we discuss the framework, the benchmarks, and the experiments conducted.

\tikzstyle{startstop} = [rectangle, rounded corners, minimum width=3cm, minimum height=1cm,text centered, draw=black]
\tikzstyle{process} = [rectangle, minimum width=3cm, minimum height=1cm, text centered, draw=black]
\tikzstyle{decision} = [diamond, minimum width=3cm, minimum height=1cm, text centered, draw=black]
\tikzstyle{arrow} = [thick,->,>=stealth]

\begin{figure}[t]
    \centering
\begin{tikzpicture}[node distance=2cm]
s
    \node[circle,fill,inner sep=1.5pt] (start) {};
    \node (preprocess) [process, below of=start] {Preprocessing};
    \node (reference) at ($(start)!0.5!(preprocess) + (2.2,0)$) {Reference Implementation};
    \node (callLLM) [process, below of=preprocess] {Interaction with the LLM};
    \node (program) at ($(preprocess)!0.5!(callLLM) + (1.6,0)$) {Prepared Program};
    \node (verification) [process, below of=callLLM] {Verification};
    \node (candidate) at ($(callLLM)!0.5!(verification) + (1.0,0)$) {Candidate};
    \node (manualFix) [right of=candidate, xshift=1cm] {Minor Fixes};

    \node (validation) [process, below of=verification] {Validation};
    \node (respond) [draw=ForestGreen, thick, circle, below of=validation, minimum size=0.5cm] {\textcolor{ForestGreen}{\checkmark}};
    
    \draw [arrow] (start) -- (preprocess);
    \draw [arrow] (preprocess) -- (callLLM);
    \draw [arrow] (callLLM) -- (verification);
    
    \node (verifyNo) [left of=verification, xshift=-0.1cm] {No};
    \node (verifyNo) at ($(verification) + (2.7,-0.3)$) {Minor issues};
    \draw [arrow, bend left, out=90, in=90] (verification.west) to (callLLM.west);
s    
    \node (verifyYes) at ($(verification)!0.5!(validation) + (0.5,0)$) {Yes};
    \node (validateYes) at ($(validation)!0.5!(respond) + (0.5,0)$) {Yes};
    \draw [arrow] (verification) -- (validation);
    \draw [arrow, densely dotted,out=0, in=270] (verification.east) to (manualFix.south);
    \draw [arrow, densely dotted,out=90, in=0] (manualFix.north) to (callLLM.east);
    
    \node (validateNo) [left of=validation, xshift=-0.4cm] {No};
    \draw [arrow, bend left, out=90, in=90] (validation.west) to (callLLM.west);
    
    \draw [arrow] (validation) -- (respond);
    
\end{tikzpicture}
    \caption{The overview of evaluating an LLM on a benchmark problem}
    \label{fig:scheme}
\end{figure}
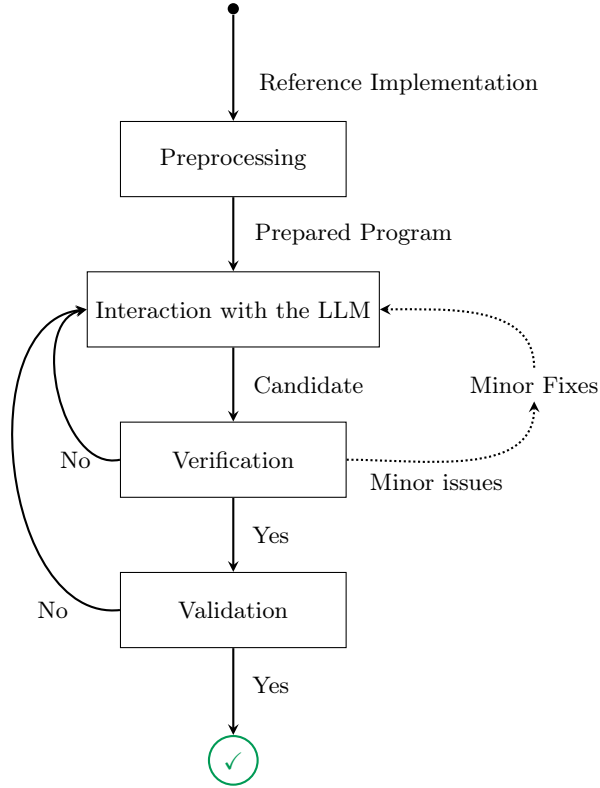

The overall pipeline consists of four key parts, see Figure~\ref{fig:scheme}. 
First, task preparation involves populating prompts with the task description and specification formed from the reference implementation in the dataset. 
Next, interaction with the large language model takes place, where the input is fed into it to generate a candidate solution. 
Following this, we attempt verification of the code produced and if any issues are detected, the feedback is passed back to the LLM to refine its suggestion. 
An optional step may be conducted to post-process the generated code with the aim of fixing common minor issues. 
Finally, we validate whether the verified code meets the reference specification. 
The last step is important to ensure that the generated preconditions are not too weak and postconditions are not too strong.  
Now, let us describe the steps in more details. 

\subsection{Task Preparation for Different Modes}

The benchmarks we use for evaluation contain ground truth, namely complete verified code that solves tasks from the original HumanEval dataset along with their natural language descriptions. 
A solution to a problem may be as small as a single function accompanied by its formal specification, if it is enough for the verifier to establish correctness. 
However, the majority of the programs include multiple functions and methods, as well as additional lemmas, loop invariants, and assertions needed for verification. 
Depending on the program synthesis task, only certain parts of a program need to be exposed. 
To simplify experiments, we annotated these components and used the annotations in the preparation of the input data. 
 


Experiment \emph{modes} describe various scenarios for which generation can be applied. 
In some of them, a model is expected to only finish the correctness proof of the existing code, while in the others it generates code from scratch. 
In different situations, we can assume either the presence of textual description of the problem or its absence. 
Naturally, the complexity of the task varies from one mode to another. 
To estimate the ability of LLMs to cope with different cases, we describe six modes, each of which requires appropriate preprocessing and a set of prompts. 

\definecolor{codegreen}{rgb}{0,0.6,0}
\definecolor{codegray}{rgb}{0.5,0.5,0.5}
\definecolor{codepurple}{rgb}{0.58,0,0.82}
\definecolor{backcolour}{rgb}{0.97,0.97,0.95}
\definecolor{forestgreen}{rgb}{0.28,0.62,0.37}
\definecolor{codeblue}{rgb}{0,0.5,1}

\lstdefinestyle{dafnystyle}{
    morekeywords={
        as, break, const, continue, crate, else, enum, extern, false, fn, for, if, impl, in, let, loop, match, mod, move, mut, pub, ref, return, self, Self, static, struct, super, trait, true, type, unsafe, use, where, while, dyn, abstract, alignof, become, box, do, final, macro, offsetof, override, priv, proc, pure, sizeof, typeof, unsized, virtual, yield, async, await, try, method, returns, then, to
    },
    sensitive=true, %
    morecomment=[l]{//},  %
    morecomment=[s]{/*}{*/},  %
    morestring=[b]",  %
    morestring=[b]{'}, %
    keywordstyle=\color{codepurple},  %
    commentstyle=\color{codegray}\itshape,  %
    stringstyle=\color{blue},  %
    identifierstyle=\color{black},  %
    ndkeywordstyle=\color{purple}\bfseries,  %
    basicstyle=\ttfamily\scriptsize,  %
    showstringspaces=false,  %
    tabsize=4,  %
    breaklines=true,  %
    breakatwhitespace=false,  %
    showtabs=false,  %
    showspaces=false,  %
    showstringspaces=false,  %
}

\lstdefinelanguage{Dafny}{
    style=dafnystyle,
    morekeywords=[2]{ requires, ensures, invariant, function, lemma, decreases, assert },
    keywordstyle=[2]\color{red}
}

\lstdefinelanguage{BigDafny}{
    basicstyle=\ttfamily,
}

\lstset{language=BigDafny}

\newcommand{\VerusBG}{\makebox[0pt][l]{\color{YellowOrange!50}\rule[-0.25em]{\linewidth}{1em}}}
\newcommand{\VerusBGlight}{\makebox[0pt][l]{\color{Yellow!50}\rule[-0.45em]{\linewidth}{1.3em}}}
\newcommand{\VerusDelete}{\makebox[0pt][l]{\color{red!20}\rule[-0.45em]{\linewidth}{1.3em}}}
\newcommand{\VerusAdd}{\makebox[0pt][l]{\color{green!20}\rule[-0.45em]{\linewidth}{1.3em}}}

\begin{figure}

\begin{minipage}[t]{0.45\linewidth}
\centering
\begin{subfigure}[t]{\linewidth}
\begin{lstlisting}[language=Dafny, escapechar=!]
function prod(s: seq<int>) : int {
  if |s| == 0 then 1 
  else s[0] * prod(s[1..])
}
method sum_product(nums: seq<int>) 
  returns (s : int, p : int)
  ensures s == sum(nums)
  ensures p == prod(nums)
{
!\VerusBG!  assert nums[..|nums|] == nums;
  s := 0, p := 1;
  for i := 0 to |nums|
!\VerusBG!    invariant s == sum(nums[..i])
!\VerusBG!    invariant p == prod(nums[..i])
    { s := s + nums[i];
      p := p * nums[i]; } ... }
  
\end{lstlisting}
    \caption{Mode 1}
    \label{fig:mode1}
\end{subfigure}
\end{minipage}
\hfill
\begin{minipage}[t]{0.45\linewidth}
\centering
\begin{subfigure}[t]{\linewidth}
\begin{lstlisting}[language=Dafny, escapechar=!]
function prod(s: seq<int>) : int {
  if |s| == 0 then 1 
  else s[0] * prod(s[1..])
}
method sum_product(nums: seq<int>) 
  returns (s : int, p : int)
!\VerusBG!  ensures s == sum(nums)
!\VerusBG!  ensures p == prod(nums)
{
!\VerusBG!  assert nums[..|nums|] == nums;
  s := 0, p := 1;
  for i := 0 to |nums|
!\VerusBG!    invariant s == sum(nums[..i])
!\VerusBG!    invariant p == prod(nums[..i])
    { s := s + nums[i];
      p := p * nums[i]; } ... }
\end{lstlisting}
    \caption{Mode 2}
    \label{fig:mode2}
\end{subfigure}
\end{minipage}

\vfill 
\begin{minipage}[t]{0.45\linewidth}
\centering
\begin{subfigure}[t]{\linewidth}
\begin{lstlisting}[language=Dafny, escapechar=!]
function prod(s: seq<int>) : int {
  if |s| == 0 then 1 
  else s[0] * prod(s[1..])
}
method sum_product(nums: seq<int>) 
  returns (s : int, p : int)
  ensures s == sum(nums)
  ensures p == prod(nums)
{
!\VerusBG!  assert nums[..|nums|] == nums;
!\VerusBG!  s := 0, p := 1;
!\VerusBG!  for i := 0 to |nums|
!\VerusBG!    invariant s == sum(nums[..i])
!\VerusBG!    invariant p == prod(nums[..i])
!\VerusBG!    { s := s + nums[i];
!\VerusBG!      p := p * nums[i]; } ... } 
\end{lstlisting}
    \caption{Mode 3}
    \label{fig:mode3}
\end{subfigure}
\end{minipage}
\hfill
\begin{minipage}[t]{0.45\linewidth}
\centering
\begin{subfigure}[t]{\linewidth}
\begin{lstlisting}[language=Dafny, escapechar=!]
function prod(s: seq<int>) : int {
  if |s| == 0 then 1 
  else s[0] * prod(s[1..])
}
method sum_product(nums: seq<int>) 
  returns (s : int, p : int)
  ensures s == sum(nums)
  ensures p == prod(nums)
{
!\VerusBG!  assert nums[..|nums|] == nums;
!\VerusBG!  s := 0, p := 1;
!\VerusBG!  for i := 0 to |nums|
!\VerusBG!    invariant s == sum(nums[..i])
!\VerusBG!    invariant p == prod(nums[..i])
!\VerusBG!    { s := s + nums[i];
!\VerusBG!      p := p * nums[i]; } ... } 
\end{lstlisting}
    \caption{Mode 4}
    \label{fig:mode4}
\end{subfigure}
\end{minipage}

\vfill 
\begin{minipage}[t]{0.45\linewidth}
\centering
\begin{subfigure}[t]{\linewidth}
\begin{lstlisting}[language=Dafny, escapechar=!]
function prod(s: seq<int>) : int {
  if |s| == 0 then 1 
  else s[0] * prod(s[1..])
}
method sum_product(nums: seq<int>) 
  returns (s : int, p : int)
!\VerusBG!  ensures s == sum(nums)
!\VerusBG!  ensures p == prod(nums)
{
!\VerusBG!  assert nums[..|nums|] == nums;
!\VerusBG!  s := 0;
!\VerusBG!  p := 1;
!\VerusBG!  for i := 0 to |nums|
!\VerusBG!    invariant s == sum(nums[..i])
!\VerusBG!    invariant p == prod(nums[..i])
!\VerusBG!    { s := s + nums[i];
!\VerusBG!      p := p * nums[i]; } ... } 
\end{lstlisting}
    \caption{Mode 5}
    \label{fig:mode5}
\end{subfigure}
\end{minipage}
\hfill
\begin{minipage}[t]{0.45\linewidth}
\centering
\begin{subfigure}[t]{\linewidth}
\begin{lstlisting}[language=Dafny, escapechar=!]
!\VerusBG!function prod(s: seq<int>) : int {
!\VerusBG!  if |s| == 0 then 1 
!\VerusBG!  else s[0] * prod(s[1..])
!\VerusBG!}
method sum_product(nums: seq<int>) 
  returns (s : int, p : int)
!\VerusBG!  ensures s == sum(nums)
!\VerusBG!  ensures p == prod(nums)
{
!\VerusBG!  assert nums[..|nums|] == nums;
!\VerusBG!  s := 0;
!\VerusBG!  p := 1;
!\VerusBG!  for i := 0 to |nums|
!\VerusBG!    invariant s == sum(nums[..i])
!\VerusBG!    invariant p == prod(nums[..i])
!\VerusBG!    { s := s + nums[i];
!\VerusBG!      p := p * nums[i]; } ... }
\end{lstlisting}
    \caption{Mode 6}
    \label{fig:mode6}
\end{subfigure}
\end{minipage}

    \caption{The \colorbox{YellowOrange!50}{dark-yellow} background highlights the parts of Dafny code required to be filled by an LLM in different modes.}
    \label{fig:modes}

\end{figure}

Figure~\ref{fig:modes} illustrates the different artifacts used in prompts distilled from the ground truth in Dafny. 
The method in the figure computes the sum and the product of the numbers in the given sequence by iterating in a for-loop. 
The function \lstinline{prod} is an example of a specification function which is not allowed to be used in the implementation. 
Notice that we have omitted some details, such as assertions, due to space constraints.
In the figure, we highlight those lines that are not included into prompts. 


\emph{Mode 1} is the least demanding scenario and represents the basic case of generating only the proof by the given specification and code. 
While preprocessing, we erase all invariants, assertions and lemma calls from the target methods. 
What is left are pre- and postconditions, specification functions, method signatures and implementations, see Figure~\ref{fig:mode1}.

\emph{Mode 2} tests generation of both a specification and a proof solely from code. 
Thus, models are given complete implementations along with specification functions, but not verification primitives. 

\emph{Mode 3} targets a situation when a developer knows exactly how a method is supposed to work and can provide the exact specification of its behavior. 
The task of an AI-assistant is to fill in the implementation and the necessary hints enabling the verifier to finish the proof. 

\emph{Mode 4} differs from Mode 3 only by requiring a text description of the program.
We believe this mode to be the most realistic and helpful use case of a tool like ours. 

The last two modes assess the capabilities of LLMs to generate everything from the implementation to a proof given only natural language specifications and method signatures. 
In \emph{Mode 5}, we provide the additional context by including specification functions in the prompt. 
However, it can be easier to guess the user intent with these functions, so we omit them in \emph{Mode 6}. 

\subsection{Interaction with an LLM}

We start interacting with an LLM by sending a system and a user prompt. 
The former primes the LLM to act as an expert in a particular verification system (Dafny, Nagini, or Verus) and defines the expected output format. 
The user prompt details the task to be performed and offers guidance on code structure, examples in a chosen language, and optional tips on relevant constructs.
Figure~\ref{fig:prompt} sketches the structure of the user prompt. 

\tikzstyle{prompt} = [rectangle, minimum width=7.5cm, minimum height=1cm, text width=6.5cm, align=left, draw=black]

\begin{figure}[t]
    \centering
\begin{tikzpicture}[node distance=1.2cm]
    \node (mode)  [prompt]                 { Mode description }; 
    \node (hints) [prompt, below of=mode]  { {\color{ForestGreen} Optional}: Language hints }; 
    \node (exmpl) [prompt, below of=hints] { {\color{ForestGreen} Optional}: Verified code sample }; 
    \node (text)  [prompt, below of=exmpl] { {\color{ForestGreen} Optional}: Textual description of the problem }; 
    \node (code)  [prompt, below of=text]  { Input code, prepared from ground truth  }; 
\end{tikzpicture}
    \caption{The general structure of a prompt}
    \label{fig:prompt}
\end{figure}
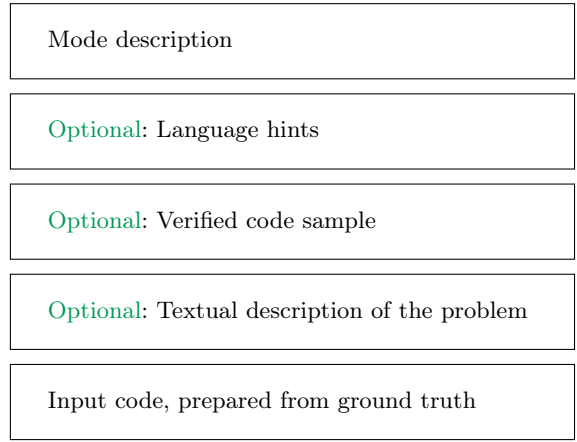

After the LLM has produced a candidate solution, its verification is attempted. 
Whenever any issues are detected in the generated code, the feedback is collected and then used in a prompt to correct them. 
This request is repeated a few times until the failure has been successfully addressed, or a limit is reached. 

We have noticed that models tend to make minor mistakes when working on Nagini, mostly mixing up keywords and syntax structures. 
For example, double negations such as \lstinline{a < b < c} are often produced even though they are not allowed in the system, likely because they are legal in Python. 
These kinds of errors can be fixed through non-ML means, which is both cheaper and faster than the counterpart. 
Thus, we implemented several simple syntactic converters to resolve such issues in Nagini and employ them prior to passing the incorrect candidate back to the LLM.

It may seem to be sufficient to stop after verifiable code is produced. 
Unfortunately, it may not be the case, as models can misinterpret user intent, break the rules by erasing function definitions or simplifying pre- and postconditions. 
Because it was our intention not to admit such poor-quality responses, we employed additional validation state. 
For scenarios that do not require producing a specification, the validation helps to check that the original code has not been modified. 
For modes aimed at generating code along with a specification, it ensures it to be sufficiently complex and expressive. 

\subsection{Validation}


When an LLM is tasked to determine formal properties that the code must satisfy, we need to ensure that generated pre- and postconditions correctly express user intentions.
Prior research either manually checks the correctness and completeness of specifications~\cite{misu2024towards} or exploits other LLMs to compare the specification and the natural language description~\cite{aggarwal2024alphaverus}.
Unfortunately, the former approach takes too much human effort, while the latter runs the risk of incorrectly assessing the quality due to the limitations of an LLM. 
Another strategy, introduced in the Clover framework~\cite{sun2024clover}, involves a formal proof of equality between annotations.  

Inspired by Clover, we check specifications by utilizing SMT-solvers.
However, instead of requiring equivalence, which may be too strict in practice, we check if the generated specification implies the specification as written in the reference solution in the data set. 
This way, we do not expect the LLM to guess the exact solution, giving it more freedom.
In particular, the generated preconditions can be weaker and the postconditions can be stronger than the original. 

To achieve this goal, we perform validation after the LLM has generated code. 
Having a list of target methods from the initial template, we construct a wrapper for each of them, that has the same specification as the original method. 
In its body, the wrapper calls the original method and propagates the return value. 
For example, for the method \lstinline{sum_product} from Figure~\ref{fig:modes}, we add a validator \lstinline{sum_product_valid} that checks the pre- and postconditions. 

Despite this approach being automatic and reliable due to its deterministic nature, it has some drawbacks. 
Firstly, there are some limitations on which language features can be used. 
For example, functions must be used instead of predicates in Dafny, classes are disallowed, as well as \lstinline{open/closed} in Verus. 
In addition to this, the validation comes with an additional verification overhead. 
Sometimes, when helpers are present, a verifier gives up trying to prove equality of the validation and the generated helpers, despite their indistinguishability. 
Finally, there might be multiple ways to specify the same user intent with sufficient accuracy. 
However, we always compare the code to the ground truth in our benchmarks, which might not scale well if more complicated tasks are considered. 

\subsection{Benchmarks}

HumanEval~\cite{chen2021evaluating} is a widely used benchmark for evaluating the code generation capabilities of LLMs.
Evaluation on HumanEval is a convenient way to showcase the results of verified code generation to a community unfamiliar with it. 
For this reason, we translated the majority of the original dataset into Dafny\footnote{HumanEval dataset in Dafny: \url{https://github.com/JetBrains-Research/HumanEval-Dafny/}} and Nagini\footnote{HumanEval dataset in Nagini: \url{https://github.com/JetBrains-Research/HumanEval-Nagini/}}, 132 and 106 programs respectively. 
We have also contributed to a similar initiative for Verus\footnote{HumanEval-based dataset in Verus: \url{https://github.com/secure-foundations/human-eval-verus}}, which contains 55 programs as of the time of writing this paper. 
Not all problems from the original benchmark have been included into the datasets, partially because of time constraints, and partially because not every task is suitable for verification. 
For some of them, the specification duplicates the implementation (e.g. task 67), while unsupported language features are needed for others (e.g. task 2 in Dafny and task 4 in Verus). 

The Dafny and Verus benchmarks were created manually, with multiple people collaborating over several weeks. 
Some automation was possible at the initial stage of creating the benchmark in Nagini. 
For this, we modified the Dafny compiler to automatically convert the Dafny files into Nagini. 
Even though the verification systems resemble each other, Nagini has lower expressiveness compared with Dafny. 
As a result, we had to manually inspect every program and add additional invariants or adjust specifications to finish correctness proofs. 

\section{Evaluation}

In the early stages of our project, we experimented with multiple LLMs, including \todo{GPT-3.5, GPT-4o, and Claude 3 Opus}. 
However, the quality achieved by Claude 3.5 Sonnet massively surpassed the others, leading us to rely exclusively on it since its release. 
Unfortunately, we have not yet run the experiments on the o1 model because of its prohibitively strict limits. 
Neither have we had a chance to set up the newest DeepSeek-R1~\cite{guo2025deepseek} and o3-mini~\cite{openAI2025o3} models, which promise even higher quality compared with the earlier models. 
In this study, we only provide the results achieved using Claude 3.5 Sonnet since it consistently demonstrated superior performance while not being too expensive in terms of time and cost.

\begin{table}[t]
    \caption{Percentage of verified examples for different modes and languages } 
    \label{tbl:overall}
    \centering
    \vspace{0.2cm}
    \begin{tabular}{l||c|r|r|r|r|r|r}
         & \begin{tabular}[c]{@{}c@{}}Number of\\ programs\end{tabular}  & \multicolumn{1}{c}{Mode 1} & \multicolumn{1}{|c|}{Mode 2} & \multicolumn{1}{c}{Mode 3} & \multicolumn{1}{|c|}{Mode 4} & \multicolumn{1}{c}{Mode 5} & \multicolumn{1}{|c}{Mode 6} \\ \hline \hline
         Dafny & 132 & 113 (86\%) & 104 (79\%) & 114 (86\%) & 108 (82\%) & 80 (61\%) & 38 (29\%) \\ 
         Nagini & 106 & 70 (66\%) & 57 (54\%) & 67 (63\%) & 67 (63\%) & 44 (42\%) & 16 (15\%) \\ 
         Verus & 55 & 25 (45\%) & 17 (31\%) & 20 (36\%) & 22 (40\%) & 13 (24\%) & 8 (15\%) \\ 
    \end{tabular}
\end{table}

In the evaluation, we allowed for five iterations to be done when trying to fix verification issues. 
The described experiment was run on HumanEval benchmarks using Claude Sonnet 3.5 five times. 
We then computed the number of unique problems, verified in at least one of the runs, and gathered the statistics in Table~\ref{tbl:overall}.

We can see that the performance of program synthesis in Dafny is higher than in either Nagini or Verus.
This is expected given that this system is more popular than the others and there is significantly more code available among the training data. 
Nevertheless, the first four modes demonstrate decent results in the case of Nagini with over half of the programs successfully verified. 
This is not the case for Verus which is the least expressive and the newest among the three. 
Note that results which are not significantly higher were achieved in the AlphaVerus project~\cite{aggarwal2024alphaverus} even though it features an advanced treefinement search and massively more calls to a model.  

The worst results were achieved in Mode 6, with less than 30\% success rate for Dafny and 15\% success rate for Nagini and Verus. 
We attribute this to our validation, which expects the specifications to match the ground truth. 

\begin{figure}
    \centering
    \begin{subfigure}[b]{0.45\textwidth}
        \centering
        \includegraphics[width=\textwidth]{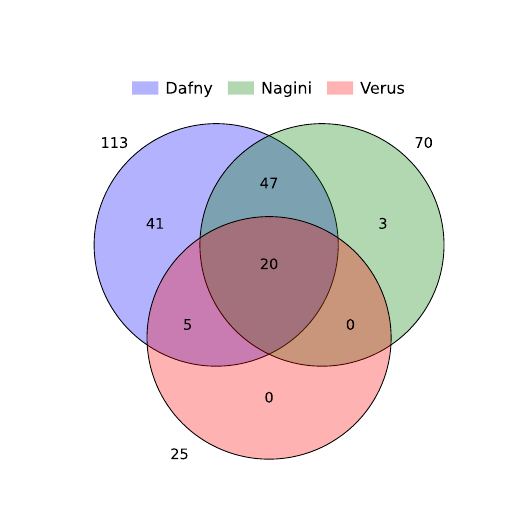}
        \caption{Mode 1}
    \end{subfigure}
    \hfill
    \begin{subfigure}[b]{0.45\textwidth}
        \centering
        \includegraphics[width=\textwidth]{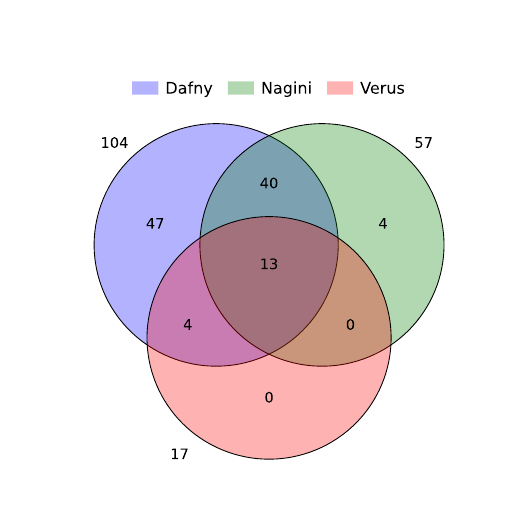}
        \caption{Mode 2}
    \end{subfigure}

    \begin{subfigure}[b]{0.45\textwidth}
        \centering
        \includegraphics[width=\textwidth]{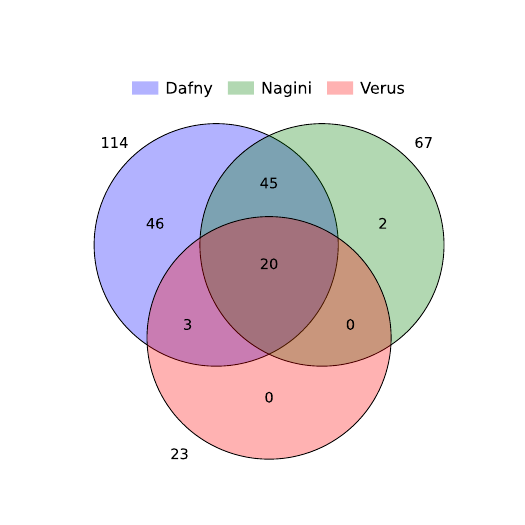}
        \caption{Mode 3}
    \end{subfigure} 
   \hfill
    \begin{subfigure}[b]{0.45\textwidth}
        \centering
        \includegraphics[width=\textwidth]{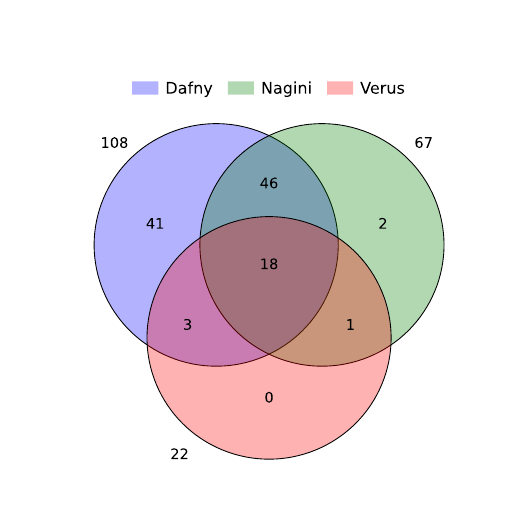}
        \caption{Mode 4}
    \end{subfigure}

    \begin{subfigure}[b]{0.45\textwidth}
        \centering
        \includegraphics[width=\textwidth]{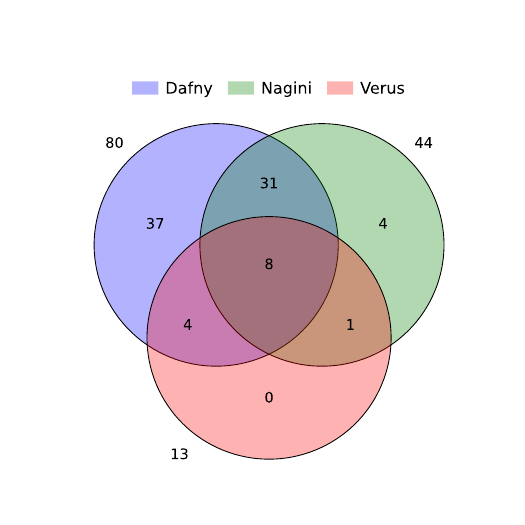}
        \caption{Mode 5}
    \end{subfigure}
    \hfill
    \begin{subfigure}[b]{0.45\textwidth}
        \centering
        \includegraphics[width=\textwidth]{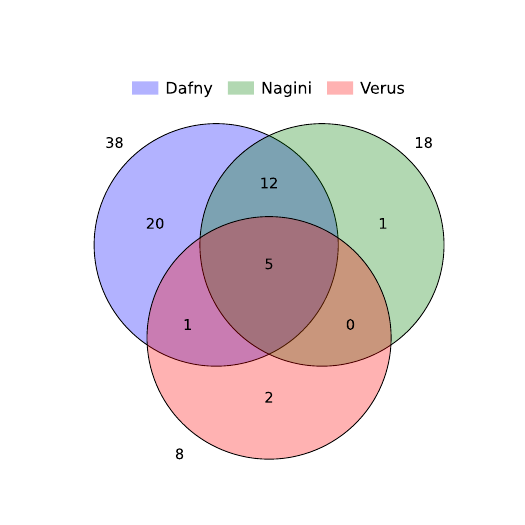}
        \caption{Mode 6}
    \end{subfigure}

    \caption{Venn Diagrams showing the intersection of unique programs successfully generated in at least one of five attempts}
    \label{fig:venn}
\end{figure}

Finally, since our benchmarks contained different subsets of tasks from the initial dataset, we also explored the distribution of problems successfully verified in each of the languages: see Figure~\ref{fig:venn}. 
Notably, only a miniscule number of problems were verified only by the systems other than Dafny. 
This may signify that the models use their knowledge of Dafny when dealing with the others.

\begin{table}[t]
    \centering
    \caption{Most common errors for different languages}

    \vspace{0.5cm}
    
    \begin{subtable}{0.45\textwidth}
        \centering
        \begin{tabular}{l||c}
             Error type & \begin{tabular}[c]{@{}c@{}}Number of\\ occurrences\end{tabular}  \\ \hline \hline
             Invariant (maintain) & 175 \\ \hline 
             Postcondition not proved & 91 \\ \hline 
             Assertion failed & 69 \\ \hline 
             Unresolved identifier & 32 \\ \hline 
             Syntax error & 27 \\ \hline 
             Invariant (entry) & 17 \\ 
        \end{tabular} 
        \caption{Dafny}
    \end{subtable}
    \hfill
    \begin{subtable}{0.45\textwidth}
        \centering
        \begin{tabular}{l||c}
             Error type & \begin{tabular}[c]{@{}c@{}}Number of\\ occurrences\end{tabular}  \\ \hline  \hline
             Timeout & 468 \\ \hline 
             Invariant (maintain) & 259 \\ \hline 
             Precondition not satisfied & 237\\ \hline 
             Postcondition not proved & 214\\ \hline 
             Invariant (entry) & 103\\  \hline 
             Unresolved identifier & 53 \\
        \end{tabular}
        \caption{Nagini}
    \end{subtable}

    \vspace{1em} 
    \begin{subtable}{0.6\textwidth}
        \centering
        \begin{tabular}{l||c}
             Error type & \begin{tabular}[c]{@{}c@{}}Number of\\ occurrences\end{tabular}  \\ \hline \hline
             Type error & 192 \\ \hline 
             Assertion failed & 46 \\ \hline 
             Invariant (maintain) & 28 \\ \hline 
             Invariant (entry) & 22\\ \hline 
             Syntax error & 17 \\ \hline 
             Arithmetic underflow/overflow & 14\\ 
        \end{tabular}
        \caption{Verus}
    \end{subtable}

\end{table}

\subsection{Understanding Common Pitfalls}

To better understand the problems that models face when generating code in different languages, we collected statistics about the errors reported by the verifier during 5 runs of the mode 1 on HumanEval benchmarks. 
We classified errors into a few groups, including syntax and type errors, unresolved identifiers, and inability to prove an invariant or a postcondition. 
Among all errors, timeout stands out: it does not occur as often in Dafny or Verus, since these languages are aimed at delivering results of verification quickly. 
Nevertheless, it is the most frequent error in the case of Nagini. 
As this error does not convey any meaningful information about the actual problem in the proof, LLMs rarely manage to resolve the issue.

While the most common errors for Dafny and Nagini are caused by incomplete proofs, Verus generation suffers from type errors. 
Incompatible types, wrong arguments, missing type annotations are most prevalent. 
Such a frequent occurrence of this type of error is due to a complex type system and the fact that different types are allowed in code and in invariants. 
As a result, the fragment of the code in Figure \ref{fig:verus:verus_code} leads to the error featured in Figure \ref{fig:verus:verus_error}. 
This error can be solved by adding explicit cast to int: \lstinline{pos as int}, but it is not often that a model is capable of it.

\begin{figure}[t]
\centering
\begin{minipage}[t]{\linewidth}
\centering
\begin{subfigure}[t]{\linewidth}
\begin{lstlisting}[language=Dafny, escapechar=!]
...
fn rolling_max(numbers: Vec<i32>) -> (result: Vec<i32>)
    ensures
        result.len() == numbers.len(),
        forall|i: int| 0 <= i < numbers.len() ==> 
            result[i] == seq_max(numbers@.take(i + 1)),
{
    let mut max_so_far = i32::MIN;
    let mut result = Vec::with_capacity(numbers.len());
    for pos in 0..numbers.len()
        invariant
            max_so_far == if pos == 0 { i32::MIN } 
                else { seq_max(numbers@.take(pos)) },
            result.len() == pos,
            forall|i: int| 0 <= i < pos ==> 
                result[i] == seq_max(numbers@.take(i + 1)),
...
\end{lstlisting}
    \caption{Implementation}
    \label{fig:verus:verus_code}
\end{subfigure}
\end{minipage}
\vfill
\begin{minipage}[t]{\linewidth}
\centering
\begin{subfigure}[t]{\linewidth}
\begin{lstlisting}[basicstyle=\ttfamily\scriptsize,  
    showstringspaces=false,  
    tabsize=4,  
    breaklines=true,  
    breakatwhitespace=false,  
    showtabs=false,  
    showspaces=false,  
    showstringspaces=false, escapechar=!]
error[E0308]: mismatched types
  --> 009-rolling_max_1.rs:24:81
   |
26 | else { seq_max(numbers@.take(pos)) },
   |                         ---- ^^^ expected `int`, found `usize`
   |                         | 
   |                         arguments to this method are incorrect
   |
\end{lstlisting}
    \caption{Type error}
    \label{fig:verus:verus_error}
\end{subfigure}
\end{minipage}

    \caption{The example of Verus code, causing the type error}
    \label{fig:verus}

\end{figure}

A more personalized approach for each types of errors could improve the quality of generated code.  
For instance, one can provide few-shot examples of typical errors and possible fixes for them. 
The Laurel project~\cite{mugnier2024laurel} intends to address the problem by modifying Dafny's error messages to include more details. 
AutoVerus~\cite{chen2024automated} proposes a rather complicated process in which unsuccessful attempts to generate proofs are collected along with the eventual correct solution to the problem. 
This information is then utilized in a self-debugging procedure, thus improving the quality considerably. 
We plan to employ these or similar methods as future work. 


\section{Conclusion}

Our study demonstrates the ability of LLMs to generate formally verified code, lowering the barriers to adoption of formal methods in mainstream programming. 
The high success rate in the context of Dafny motivates further research into less-explored verification systems, such as Nagini and Verus. 
Mistakes that LLMs tend to make for these systems likely stem from the models' unfamiliarity with them, which we plan to address in future work by fine-tuning. 
This will require significantly larger datasets, the collection of which is complicated by the insufficient amount of source code published online, but can be approached through synthetic means. 
Another research direction can leverage specialized error-correction mechanisms and self-improving frameworks. 
Finally, striking the right balance between automation and human involvement can be key to making verification accessible to the wider programming community.

\bibliographystyle{plain}
\bibliography{main}

\end{document}